\documentstyle[12pt]{article}

\begin{document}
{\sf \begin{center} \noindent
{\Large \bf The role of curvature and stretching on the existence of fast dynamo plasma in Riemannian space}\\[3mm]

by \\[0.3cm]

{\sl L.C. Garcia de Andrade}\\

\vspace{0.5cm} Departamento de F\'{\i}sica
Te\'orica -- IF -- Universidade do Estado do Rio de Janeiro-UERJ\\[-3mm]
Rua S\~ao Francisco Xavier, 524\\[-3mm]
Cep 20550-003, Maracan\~a, Rio de Janeiro, RJ, Brasil\\[-3mm]
Electronic mail address: garcia@dft.if.uerj.br\\[-3mm]
\vspace{2cm} {\bf Abstract}
\end{center}
\paragraph*{}
\textbf{Vishik's antidynamo theorem is applied to non-stretched
twisted magnetic flux tube in Riemannian space. Marginal or slow
dynamos along curved (folded), torsioned (twisted) and
non-stretching flux tubes plasma flows are obtained}. Riemannian
curvature of twisted magnetic flux tube is computed in terms of the
Frenet curvature in the thin tube limit. It is shown that, for
non-stretched filaments fast dynamo action in diffusive case cannot
be obtained, in agreement with Vishik's argument, that fast dynamo
cannot be obtained in non-stretched flows. \textbf{In this case a
non-uniform stretching slow dynamo is obtained}.\textbf{An example
is given which generalizes plasma dynamo laminar flows, recently
presented by Wang et al [Phys Plasmas (2002)], in the case of low
magnetic Reynolds number $Re_{m}\ge{210}$. Curved and twisting
Riemannian heliotrons, where non-dynamo modes are found even when
stretching is presented, shows that the simple presence of
stretching is not enough for the existence of dynamo action. Folding
is equivalent to Riemann curvature and can be used to cancell
magnetic fields, not enhancing the dynamo action. In this case
non-dynamo modes are found for certain values of torsion or Frenet
curvature (folding) in the spirit of anti-dynamo theorem. It is
shown that curvature and stretching are fundamental for the
existence of fast dynamos in plasmas.}{\bf PACS
numbers:\hfill\parbox[t]{13.5cm}{02.40.Hw:differential geometries.
91.25.Cw-dynamo theories.}}

\newpage
\newpage
 \section{Introduction}
 \textbf{An anti-dynamo theorem by Zeldovich \cite{1}, showns that planar flows cannot support
 dynamo action. Together with Cowling's theorem \cite{2}, \textbf{which considers that axisymmetric magnetic fields cannot give rise to dynamo
 action, they form the more traditional and well-tested antidynamo
 theorems ever}. Actually, one of the main ingredients in dynamo
 action is exponential stretching. This Lyapunov exponential stretching was previously investigated by Friedlander and Vishik
 \cite{3}. \textbf{Because of the cancellation of magnetic fields in
 regions of strong folding or
 curvature, presence of stretching alone, is not enough to warrant fast dynamo action. This is mainly due to the fact that, cancellation of
 magnetic fields are possible in regions of folding or curvature}. In tubes, the Riemann curvature could naturally cause these kind of problems. However,
 one may always consider that inside the tubes magnetic fields are
 so concentrated that if they do not strongly curves of folds, this
 cancellation cannot be effective. Thus Riemann curvature would have to be too
 strong in order that cancellation takes place}.
 Friedlander and Vishik ,argued that there
 are several topological obstructions to the existence of Anosov flows \cite{4} in Riemannian 3-D space. Anosov
 flows, \textbf{are Riemannian spaces of constant negative curvature, endowed with
 geodesic flows,} have \textbf{also} been \textbf{considered} by Chicone
 and Latushkin \cite{5} \textbf{who showed} that they are simple \textbf{example of} fast dynamos in compact Riemannian manifolds.
 Other types of fast dynamo mechanisms \textbf{based on}
 stretching flux tubes in Riemannian conformal manifolds, have been obtained recently by
 Garcia de Andrade \cite{6}. Yet much earlier, M. Vishik \cite{7}
 has argued that only slow dynamos can be obtained from non-stretching dynamo flows. \textbf{Actually no fast dynamos obtained by Vainshtein and
 Zeldovich \cite{8} stretch-twist and fold (STF) method \cite{9}
 can be obtained without stretching}. This sort of "anti-fast-dynamo" theorem on flows, can here
 be \textbf{generalized} to flux tube dynamos and filaments. \textbf{In} this paper it is shown that the Vishik argument can be extended
 to \textbf{produce} \textbf{an} anti-fast-dynamo theorem for filaments and tubes in some particular cases, subject to bounds in poloidal and
 toroidal magnetic fields, \textbf{as well as to} the twist (torsion) of the magnetic flux tube axis. Exponential stretching of
 tube is \textbf{derived}. \textbf{Bounds on the growth rate of fast dynamo action, based on line-stretching, have been given by Klapper and Young
 \cite{10}. Non-stretching in diffusionless media is used for flux
 tube dynamos}, while diffusive filaments are used in the second case.
 Marginal dynamos are obtained in steady dynamos,
 \textbf{where} constraints are placed on the magnetic fields \textbf{in order} dynamo action \textbf{can} be effective. Diffusion processes,
 have been previously investigated by S. Molchanov \cite{11}, \textbf{also} in the context of Riemannian geometry.
 Such slow dynamos have been obtained by Soward \cite{12}, which
 argued that fast dynamo actions would \textbf{still} be possible in regions where no
 non-stretching flows would be presented,such as in some curved
 surfaces. \textbf{It is clear, from this paper,} that these surfaces could be of Riemannian. In filaments , it is shown that
 dynamo action in non-stretched unfolded and untwisting filaments could be obtained.
 When torsion vanishes, filaments are planar, which by Zeldovich
 anti-dynamo theorem \cite{1} imply that dynamo action cannot be supported. Actually, it is shown that it
 generates a static magnetic initial field and a steady perturbation which may be, at best, a marginal dynamo.
 To \textbf{resume}, fast dynamos are generated by stretching, folding
 and twisting of the loops or filaments. It seems that non-stretched
 ,folded and twisting filaments leads to slow dynamos filaments. \textbf{Earlier, Mikhailovskii \cite{13} has presented several Riemannian 3D metrics to
 describe various types of plasma devices such as heliotrons, tokamaks
 and stellarators. One year earlier Riemannian geometry was used by Ricca \cite{14} to investigate inflexional instabilities in solar
 physics plasma loop. More recently, Ricca's geometry has been applied to curved twisted flux tube with current
 carrying loops \cite{15}, helical plasmas and conformal fast dynamos \cite{6}. Another interesting and important application of Riemannian geometry
 to dynamo theory was given by Arnold and Arnold et al \cite{16} in a somewhat artificial uniform exponential stretching to produce a chaotic fast dynamo.
 In 2002 a new kind of dynamo based on flowing laminar laboratory
 plasmas has been obtained by Wang et al \cite{17}. Its plasma kinetic
 to magnetic energy conversion has been verified by numerical
 kinematical simulations. Their plasma flow topology is meant to substitute the liquid-sodium intrinsically turbulent dynamos \cite{17}.
 Magnetic field free
 coaxial plasma guns \cite{18} can be used to sustain this plasma dynamo flow. Since analytical solutions of magnetic vector potential are very hard
 to find in diffusive media, in this letter one considers the generalization of these plasma dynamo laminar flows to the curved and twisting flow of
 heliotrons to arbitrary magnetic Reynolds numbers by solving the equations one for $Re_{m}\rightarrow{\infty}$ approximation. Though this approximation
 is more suitable to astrophysical applications than to plasma
 terrestrial LAB devices, the simplication
 and the results obtained seems to pay off on a first approach to the problem. By tuning the physical heliotron parameters such as magnetic vector potential
 components and twisting one obtains a constraint between the Frenet constant torsion and the number of revolutions along the toroidal direction and by assuming that the
 growth factor ${\gamma}$ of magnetic field, a non-dynamo mode is obtaining under these constraint. The
 curved and twisted Riemannian helical tube is thick as in the case of laminar plasma dynamo \cite{19}. The paper is organized as
 follows: In section 2 a review on dynamics of holonomic Frenet frame is presented along with the discussion of
 vortex-filament exponential stretching. In section
 3 the self-induction equation is solved in the framework of the Ricca's \cite{14} twisted magnetic flux tube in Riemannian
 3D manifold. In this same section the "new" anti-dynamo theorem is presented. In section 4 a
 twisted (torsioned) curved filament is shown to be a fast
 dynamo in Euclidean space. Section 5 contains the plasma non-dynamo modes example. In section 6, another, more realistic example of a stretched, non-dynamo
 flow is obtained in non-stretching regions. The converse of this case , where the fast dynamo can be obtained in stretching regions in the
 middle of non-stretching ones, is exactly the expression of Soward's conjecture given
 above. Discussions and future prospects are presented in section 7}.
\newpage
\section{Exponential stretching in dynamo flux tube} This section contains a brief review
of the Serret-Frenet holonomic frame \cite{20} equations that are
specially useful in the investigation of STF Riemannian flux tubes
in magnetohydrodynamics (MHD) with magnetic diffusion. \textbf{Here
the Frenet frame is attached along the magnetic flux tube axis which
possesses Frenet torsion ${\tau}(s)$ and curvature ${\kappa}(s)$,
which completely determine topologically the filaments}. Some
dynamical relations from vector analysis and differential geometry
of curves for the Frenet frame $(\textbf{t},\textbf{n},\textbf{b})$
are given by
\begin{equation}
\textbf{t}'=\kappa\textbf{n} \label{1}
\end{equation}
\begin{equation}
\textbf{n}'=-\kappa\textbf{t}+ {\tau}\textbf{b} \label{2}
\end{equation}
\begin{equation}
\textbf{b}'=-{\tau}\textbf{n} \label{3}
\end{equation}
\textbf{Here $\textbf{t}$ is the tangent vector to the magnetic
toroidal component, while $\textbf{n}$ and $\textbf{b}$ lie on an
orthogonal plane to vector $\textbf{t}$}. The holonomic dynamical
time evolution relations are obtained from vector analysis and
differential geometry of curves as
\begin{equation}
\dot{\textbf{t}}=[{\kappa}'\textbf{b}-{\kappa}{\tau}\textbf{n}]
\label{4}
\end{equation}
\begin{equation}
\dot{\textbf{n}}={\kappa}\tau\textbf{t} \label{5}
\end{equation}
\begin{equation}
\dot{\textbf{b}}=-{\kappa}' \textbf{t} \label{6}
\end{equation}
Together with the flow derivative
\begin{equation}
\dot{\textbf{t}}={\partial}_{t}\textbf{t}+(\vec{v}.{\nabla})\textbf{t}
\label{7}
\end{equation}
From these equations the generic flow \cite{13}
\begin{equation}
\dot{\textbf{X}}=v_{s}\textbf{t}+v_{n}\textbf{n}+v_{b}\textbf{b}
\label{8}
\end{equation}
\textbf{leads to} the line-stretching equation
\begin{equation}
\frac{{\partial}l}{{\partial}t}=(-\kappa{v}_{n}+{v_{s}}')l\label{9}
\end{equation}
where $l$ is given by
\begin{equation}
l:=(\textbf{X}'.\textbf{X}')^{\frac{1}{2}}\label{10}
\end{equation}
Solution of equation (\ref{9}) \textbf{results in}
\begin{equation}
l=l_{0}e^{\int{(-\kappa{v}_{n}+{v_{s}}')dt}}\label{11}
\end{equation}
\textbf{this} shows that when the component along the tangent vector
$\textbf{t}$, $v_{s}=v_{0}$ is constant, the solenoidal
incompressible flow \textbf{equation reads}
\begin{equation}
{\nabla}.\textbf{v}=0=({v'}_{s}-{\kappa}v_{n})\label{12}
\end{equation}
This implies that \textbf{if} $v_{n}$ vanishes, one is left with a
non-stretched twisted flux tube flow. This choice
$\textbf{v}=v_{0}\textbf{t}$, where $v_{0}=constant$ is the steady
flow \textbf{choice} one uses here. This definition of magnetic
filaments is obtained from the solenoidal \textbf{character} of the
magnetic field
\begin{equation}
{\nabla}.\textbf{B}=0\label{13}
\end{equation}
where $B_{s}$ is \textbf{its} toroidal component. In the next
section one \textbf{solves} the diffusion equation in the steady
case. \textbf{In} the non-holonomic Frenet frame \textbf{such
equation is written as}
\begin{equation}
{\partial}_{t}\textbf{B}={\nabla}{\times}(\textbf{v}{\times}\textbf{B})+{\eta}{\nabla}^{2}\textbf{B}
\label{14}
\end{equation}
Here ${\eta}$ \textbf{represents} the magnetic diffusion. Since in
astrophysical scales,
${\eta}{\nabla}^{2}\approx{{\eta}{L^{-2}}}\approx{{\eta}{\times}10^{-20}}cm^{-2}$,
for a solar loop scale length of $10^{10}cm$ \cite{21} one notes
that the diffusion effects can be neglected. Let us now consider the
magnetic field definition in terms of the magnetic vector potential
$\textbf{A}$ as
\begin{equation}
\textbf{B}={\nabla}{\times}\textbf{A} \label{15}
\end{equation}
where the gradient operator is
\begin{equation}
{\nabla}=\textbf{t}{\partial}_{s}+\frac{1}{r}\textbf{e}_{\theta}+\textbf{e}_{r}{\partial}_{r}\label{16}
\end{equation}
\newpage
Let us now consider the Riemann metric of a flux tube in curvilinear
coordinates $(r,{\theta}_{R},s)$, where
${\theta}={\theta}_{R}-\int{{\tau}(s)ds}$. \textbf{The integral of
torsion along the filaments is called total torsion}. This metric is
encoded into the Riemann line element
\begin{equation}
d{s_{0}}^{2}= dr^{2}+r^{2}d{{\theta}}^{2}+K^{2}ds^{2} \label{17}
\end{equation}
where, accordingly to our hypothesis,
$K^{2}=(1-{\kappa}rcos{\theta})^{2}$. \textbf{This expression}
contributes to the Riemann curvature components. These can be easily
computed with the tensor \textbf{computer} package in terms of
Riemann components \textbf{input}. \textbf{Adapting} the general
relativity (GR) tensor package to three-dimensions, yields
\textbf{the following curvature components expressions}
\begin{equation}
R_{1313}=R_{rsrs}=
-\frac{1}{4K^{2}}[2K^{2}{\partial}_{r}A(r,s)-A^{2}]=-\frac{1}{2}\frac{K^{4}}{r^{2}}=-\frac{1}{2}r^{2}{\kappa}^{4}cos^{2}{\theta}
\label{18}
\end{equation}
\begin{equation}
R_{2323}=R_{{\theta}s{\theta}s}= -\frac{r}{2}A(r,s)=
-{K^{2}}\label{19}
\end{equation}
where $A:={\partial}_{r}K^{2}$. \textbf{For} thin tubes,
$K^{2}(r,s)\approx{1}$ $(r\approx{0})$, \textbf{and} these Riemann
curvature components reduce to
\begin{equation}
R_{1313}=R_{rsrs}= -\frac{1}{r^{2}} \label{20}
\end{equation}
\textbf{Thus} Riemann curvature of the tube is particularly strong
\textbf{as} the tube axis ($r=0$) \textbf{is approached}. In the
next section some of the mathematical machinery derived
\textbf{here} \textbf{is used in} the formulation of a new
anti-dynamo theorem. \textbf{Let us now consider the generic flow in
flux tube Riemannian space}. As in Ricca's \cite{14}, \textbf{here}
one considers that no radial components of either flows or magnetic
fields are present. \textbf{These assumptions} yields
\begin{equation}
\dot{\textbf{X}}=v_{s}\textbf{t}+v_{\theta}\textbf{e}_ {\theta}
\label{21}
\end{equation}
\textbf{By considering the relation between the base
$(\textbf{e}_{r},\textbf{e}_{\theta},\textbf{t})$ and Frenet frame}
\begin{equation}
\textbf{e}_{r}=cos{\theta}\textbf{n}+sin{\theta}\textbf{b}\label{22}
\end{equation}
and
\begin{equation}
\textbf{e}_{\theta}=-sin{\theta}\textbf{n}+cos{\theta}\textbf{b}\label{23}
\end{equation}
one may express formula (\ref{21}) as
\begin{equation}
\dot{\textbf{X}}=v_{s}\textbf{t}-v_{\theta}sin{\theta}\textbf{n}+v_{\theta}cos{\theta}\textbf{b}
\label{24}
\end{equation}
Comparison between (\ref{24}) and (\ref{21}) yields
\begin{equation}
v_{n}=-v_{\theta}sin{\theta} \label{25}
\end{equation}
and
\begin{equation}
v_{b}=v_{\theta}cos{\theta} \label{26}
\end{equation}
which by a simple comparison with expression (\ref{22}) yields
\begin{equation}
\frac{{\partial}l}{{\partial}t}=(\kappa{v}_{\theta}sin{\theta}
+{v_{s}}')l\label{27}
\end{equation}
\textbf{Therefore} one finally finds an expression \textbf{for} the
exponential stretching l as
\begin{equation}
l=l_{0}e^{\int{(\kappa{v}_{\theta}sin{\theta}
+{v_{s}}')dt}}\label{28}
\end{equation}
From the solenoidal incompressible flow equation
\begin{equation}
{\nabla}.\textbf{v}=0\label{29}
\end{equation}
one obtains
\begin{equation}
{\partial}_{s}{v}_{\theta}=r{\kappa}{\tau}sin{\theta}{v_{\theta}}\label{30}
\end{equation}
where the operator ${\partial}_{\theta}=-{\tau}^{-1}{\partial}_{s}$,
and flux tube definition \textbf{of} twist angle
${\theta}={\theta}_{R}-{\int{{\tau}ds}}$ \textbf{were used in the
computation}. \textbf{Substituting} this result into expression
(\ref{28}), \textbf{and assuming as in Ricca's paper that tube
${v'}_{s}$ vanishes}, integration yields
\begin{equation}
l=l_{0}e^{(\frac{1}{{\tau}_{0}a}\int{{v'}_{\theta}dt})} \label{31}
\end{equation}
where filament cross-section \textbf{is assumed to be constant}.
\textbf{To simplify computations it is assumed that the dynamos are
helical, which means that the torsion and Frenet curvature are taken
as equal and constants $(r=a)$}. By changing the integrand according
to ${{v'}_{\theta}}dt=\frac{{v'}_{\theta}}{v_{s}}$ where
$dt={v_{s}}^{-1}ds$ , formula (\ref{31}) \textbf{reads}
\begin{equation}
l=l_{0}e^{(\frac{v_{\theta}}{v_{0}{\tau}_{0}a}) }\label{32}
\end{equation}
where $v_{s}\approx{v_{0}}$ is taken to simplify matters. This shows
that the torsion of the tube decreases the tube axis \textbf{length}
and the poloidal flow enhances tube's twist. \textbf{This is similar
to the piece of cloth twist example , where when the piece of cloth
is twisted it ia also stretched. This result is actually confirmed
by Klapper and Longcope \cite{22} assertion that the twist is
influenced by the flow where tubes are immersed. Actually the
stretching of the tube and its torsion are related by the above
expression (\ref{11}), and torsion appears in the exponential
stretching since in helical dynamos torsion equals the Frenet
curvature $\kappa$. Now let us turn our attention to the plasma
rotation inside the flux tube which can be obtained from the
vorticity equation}
\begin{equation}
{\nabla}.\vec{{{\omega}}}={\nabla}.{\nabla}{\times}\textbf{v}=0\label{33}
\end{equation}

\newpage

which yields the following PDEs
\begin{equation}
{\omega}_{r}=-{\partial}_{s}v_{\theta}\label{34}
\end{equation}
\begin{equation}
{\omega}_{\theta}={\omega}_{0}=-{\partial}_{r}v_{s}\label{35}
\end{equation}
\begin{equation}
{\omega}_{s}=-[{\partial}_{r}v_{\theta}-\frac{cos{\theta}}{r}{\tau}_{0}v_{\theta}]\label{36}
\end{equation}
From vorticity expression (\ref{35}) one obtains
\begin{equation}
{\omega}_{0}r=-v_{s}\label{37}
\end{equation}
Substitution of the expression for ${\partial}_{s}v_{\theta}$ above
into expression (\ref{34}) now yields
\begin{equation}
{\omega}_{r}=-{{\tau}_{0}}^{2}rsin{\theta}v_{\theta}\label{38}
\end{equation}
for the thin flux tube. For steady dynamos one obtains \cite{17}
\begin{equation}
(\textbf{v}.{\nabla})\textbf{B}=(\textbf{B}.{\nabla})\textbf{v}\label{39}
\end{equation}

\begin{equation}
\frac{B_{\theta}}{B_{s}}=\frac{v_{\theta}}{v_{s}}\label{40}
\end{equation}
\textbf{By making use of the vorticity equations one obtains}
\begin{equation}
\frac{B_{s}}{B_{\theta}}\approx{{\tau}_{0}rcos{\theta}}\label{41}
\end{equation}
Note then that near to the flux tube axis $(r\approx{0})$ the
toroidal magnetic field \textbf{decreases with respect to the from
toroidal field}. This expression is similar to one obtained by Ricca
in terms of twist, and \textbf{it seems to be important} for plasma
fusion devices, where magnetic field lines twisting can be adjusted
\textbf{in order turbulent damping effects takes place}. A dynamo
test can be obtained by computing the integral of Zeldovich
\begin{equation}
\frac{4{\pi}d{\epsilon}_{M}}{dt}=\int{\textbf{B}.(\textbf{B}.{\nabla})\textbf{v}dV}
\label{42}
\end{equation}
where another \textbf{diffusion term has been dropped}. This
expression can be applied to the flux tube if one considers that the
above integrand is computed in detail as ( \textbf{Note to the
referee: This equation has been fixed and detailed as demanded in
one of yu kind querries! Thanks!}
\begin{equation}
(\textbf{B}.{\nabla})\textbf{B}=[B_{s}-\frac{{B}_{\theta}{{\tau}_{0}}^{-1}}{r}]{\partial}_{s}(v_{\theta}\textbf{e}_{\theta}+v_{s}\textbf{t})
\label{43}
\end{equation}
which when \textbf{scalarly multiplied by} $\textbf{B}$
yields(\textbf{note to the referee: this equation has also been
corrected}}
\begin{equation}
\textbf{B}.[(\textbf{B}.{\nabla})\textbf{B}]=[-B_{s}v_{\theta}{\tau}_{0}+{B}_{\theta}{{\tau}_{0}}^{2}(v_{\theta}-v_{s}{{\tau}_{0}}^{-1})]
\label{44}
\end{equation}
which finally yields
\begin{equation}
\frac{4{\pi}d{\epsilon}_{M}}{dt}=\int{[{B}_{\theta}{{\tau}_{0}}^{2}sin{\theta}[v_{\theta}-{{\tau}_{0}}^{-1}v_{s}]+
B_{s}v_{\theta}](B_{s}-\frac{B_{\theta}{{\tau}_{0}}^{-1}}{r})dV}
\label{45}
\end{equation}
which \textbf{shows} that
\begin{equation}
B_{s}=\frac{B_{\theta}{{\tau}_{0}}^{-1}}{r} \label{46}
\end{equation}
implies the existence of a marginal dynamo, where $
\frac{4{\pi}d{\epsilon}_{M}}{dt}$ vanishes. Note that this result
does not depend directly on the stretching but on the torsion which
\textbf{is} indirectly responsible for stretching. \textbf{Therefore
exponential stretching may exists even for marginal dynamos}, which
would represent a converse result of a Vishik's antidynamo theorem
for flux tubes. Non-stretching flows implies \textbf{necessarily the
existence of slow dynamos}. \textbf{Nevertheless the converse is not
true}, slow dynamos do not necessarily \textbf{imply}
non-stretching. The weak torsion approximation used here,is in
agreement with the weak torsion (twist) \textbf{found} in the solar
twisted coronal loop torsion $({\tau}_{0}\approx{10^{-10}cm^{-1}})$
\cite{21}.
\newpage

\section{Anti-dynamo theorem in non-stretching filaments}
In this section the anti-dynamo formulation of non-stretching dynamo
flows in twisted filaments is presented. This can be done by simply
considering the gradient along the filament as
${\nabla}=\textbf{t}{\partial}_{s}$ and computing the total magnetic
energy on a diffusive medium as
\begin{equation}
\frac{4{\pi}d{\epsilon}_{M}}{dt}=\int{{B_{s}}^{2}v_{s}\textbf{t}.\textbf{n}dV}:=0
\label{47}
\end{equation}
\textbf{since $\textbf{n}.\textbf{t}=0$} (\textbf{note to the
referee: This clearly shows that $\textbf{b}$ is the other leg in
the frame and not the magnetic field as previously thought by the
referee)}. Here one has considered that
$\textbf{B}:=B_{s}\textbf{t}$. Thus one can say that in a diffusive
media, when ${v'}_{s}=0$ and $v_{n}$ \textbf{both} vanishes the
following lemma is proved.

\textbf{lemma}:

Non-stretching vortex filaments in a diffusionless media gives rise
to a marginal or slow dynamo. No fast dynamo being possible.

This can be considered as a sort of anti-fast dynamo theorem
motivated from Vishik's idea. Now let us introduce diffusion into
the problem. As previously \textbf{noted} by Zeldovich \cite{1},
this leads us to the dynamo action and possibly to fast dynamos. In
the case of diffusive filaments one notes that the magnetic
induction equation
\begin{equation}
\frac{d}{dt}\textbf{B}=(\textbf{B}.{\nabla})\textbf{v}+{\eta}{\nabla}^{2}\textbf{B}
\label{48}
\end{equation}
where ${\eta}$, the magnetic resistivity or diffusion, is considered
as constant, \textbf{reduces} the problem to the three scalar
equations
\begin{equation} \frac{d{B}_{s}}{dt}=-{\eta}{\kappa}{B}_{s} \label{49}
\end{equation}
\begin{equation} {\kappa}'={\eta}{\kappa}\tau \label{50}
\end{equation}
\begin{equation} -\kappa{\tau}={\eta}{\kappa}'+v_{s} \label{51}
\end{equation}
Solution of these equations \textbf{yields}
\begin{equation} {B}_{s}=exp[-{\eta}\int{{\kappa}^{2}ds}]{B}_{0} \label{52}
\end{equation}
\textbf{where $\int{{\kappa}^{2}ds}$ is the total Frenet curvature
energy integral}. The remaining solutions are
\begin{equation} {\kappa}=exp[{\eta}\int{{\tau}ds}]{\kappa}_{0} \label{53}
\end{equation}
where $\int{{\tau}ds}$ is the total torsion. \textbf{In} the case of
helical filaments, where torsion equals curvature,\textbf{one
obtains}
\begin{equation} -{{\tau}_{0}}^{2}=v_{s} \label{54}
\end{equation}
Note that the decaying of the magnetic field depends on the sign of
the integral, but since the average value of this integral is
positive, the magnetic field decays and no dynamo action is
possible, \textbf{and} even slow dynamos cannot be found.
\textbf{In} the case of non-stretching filaments, the presence of
diffusion actually enhances the non-dynamo \textbf{character} of
Vishik's lemma.
\section{Filamentary fast dynamos in Euclidean space}
\textbf{Earlier, Arnold et al \cite{16} found a stretching and
compressed fast dynamo in curved Riemannian space. Though this
served as motivation for the study of flux tube dynamos in 3D curved
Riemannian space, in this section one shall address the problem of
finding a filamentary fast dynamo in Euclidean space in 3D}. The
stretching followed by squeezing is a path to finding a growing
magnetic field. Recently Nu\~{n}ez \cite{22} has considered a
similar problem , also making use of Frenet frame as here, by
investigating eigenvalues in the stretching plasma flows. In this
section one shows that the stretching condition in filaments is
fundamentally connected to the incompressibilty of the flow. This is
simply understood if one considers the exponent in the stretching in
expression
\begin{equation}
{\gamma}:=-\kappa{v}_{n}+{v_{s}}'\label{55}
\end{equation}
\textbf{which shows that stretching factor gamma is a fundamental
quantity to be examined when one wants to find out a fast dynamo
action}. Note that when ${\gamma}\ge{0}$ or ${\gamma}<0$, one would
have respectively either a fast or slow or marginal dynamo and a
decaying magnetic field as found before. Let us now drop the
constraint that $v_{s}$ is constant and substitute the flow
\begin{equation}
\textbf{v}={v}_{n}\textbf{n}+{v_{s}}\textbf{t}\label{56}
\end{equation}
into the solenoidal incompressible flow
\begin{equation}
{\nabla}.\textbf{v}=0\label{57}
\end{equation}
\textbf{with vanishing} $v_{n}$, one should have \textbf{a}
non-stretched twisted flux tube. This is exactly the choice
$\textbf{v}=v_{0}\textbf{t}$, where $v_{0}=constant$ is the steady
flow one uses here. This definition of magnetic filaments is
\textbf{obtained} from the solenoidal \textbf{character} of the
magnetic field
\begin{equation}
{\nabla}.\textbf{v}=-\kappa{v}_{n}+{v_{s}}'=0\label{58}
\end{equation}
This result implies that ${\gamma}=0$ which \textbf{in turn} implies
no flow stretching at all! This \textbf{leads} us to note that if a
fast filamentary dynamo action \textbf{is} possible, a modification
of the flow has to be performed. To investigate this possibility one
considers the following form of the dynamo flow
\begin{equation}
\textbf{v}={v}_{n}\textbf{n}+{v_{s}}\textbf{t}+v_{0}\textbf{b}\label{59}
\end{equation}
which minimally generalizes (\ref{56}). In this case, the above
expression for ${\gamma}$ does not vanish and is equal to
\begin{equation}
{\gamma}=-\kappa{v}_{n}+{v_{s}}'={\tau}_{0}v_{0}\label{60}
\end{equation}
Note that again for ${\gamma}>0$ one obtains a fast dynamo since
${\eta}=0$ and stretching is possible if ${\tau}_{0}$ and $v_{0}$
possesses the same sign. Actually this flow leads to the following
three scalar dynamo equations
\begin{equation}
\frac{d}{dt}{B}_{s}={\gamma}{B}_{s}-{{\tau}_{0}}^{2}B_{n} \label{61}
\end{equation}
\begin{equation}
\frac{d}{dt}{B}_{n}={\tau}_{0}(v_{s}-{\tau}_{0}-\frac{1}{{\tau}_{0}}{\partial}_{s}v_{n}){B}_{s}
\label{62}
\end{equation}
\begin{equation}
{B}_{s}{{\tau}_{0}}v_{n}=0 \label{63}
\end{equation}
Here ${\gamma}=({v_{s}}'-{\tau}_{0}v_{n})$ since we are considering
helical dynamo filaments. From equation (\ref{62}) one obtains
$v_{n}=0$ which \textbf{further} simplifies the other equations.
Since in astrophysical scales , torsion is as weak as
${{\tau}_{0}}^{2}\approx{{\eta}{\times}10^{-20}}cm^{-2}$,
\textbf{for a solar coronal loop scale, the terms proportional to
torsion squared may be dropped}. In this approximation a fast dynamo
solution is found from (\ref{61}) as (\textbf{note to the referee:
these equations are new)\begin{equation} {B}_{s}=B_{0}e^{{\gamma}t}
\label{64}
\end{equation}
and
\begin{equation}
\frac{d}{dt}{B}_{n}={\tau}_{0}(v_{s}-{\tau}_{0}){B}_{s} \label{65}
\end{equation}
which yields
\begin{equation}
{B}_{n}=\frac{1}{{v}_{0}}(v_{s}-{\tau}_{0})e^{{\gamma}t} \label{66}
\end{equation}}
where to obtain $B_{n}(t,s)$ use \textbf{has been made} of the
stationary property of the flow ${\partial}_{t}v_{s}=0$, in order
\textbf{of} being able to integrate (\ref{64}). Thus a fast dynamo
action for filamentary flows in Euclidean 3D space \textbf{has been
obtained}. Note that this action is natural and \textbf{actually}
somewhat expected, since when one changes the dynamo
three-dimensional $(v_{s},v_{n},v_{b})$ a two dimensional flow
$(v_{s},v_{n})$ is obtained , which is strictly forbbiden from
Zeldovich anti-dynamo theorem. Besides if one integrates the flow
equation (\ref{59}) yields (\textbf{note to the referee: This
formula is new})
\begin{equation}
{v_{s}}={\tau}_{0}v_{0}s+c_{1}\label{67}
\end{equation}
where $c_{1}$ is an integration constant. Expression (\ref{67})
indicates that the flow along the filament undergoes a uniform
strain \cite{18} due to the stretching of the magnetic filament. Let
us now consider the approximation of weak torsion given in solar
plasma loops for example. In this approximation the above relations
reduce to
\begin{equation}
\frac{d}{dt}{B}_{n}={\tau}_{0}v_{s}{B}_{s} \label{68}
\end{equation}
which yields
\begin{equation}
{B}_{n}=\frac{{\tau}_{0}c_{1}B_{0}}{\gamma}e^{{\gamma}t} \label{69}
\end{equation}
Now if one uses these relations into the total magnetic energy
integral, one obtains
\begin{equation}
{\epsilon}_{M}=
\frac{a^{2}}{8}[{B_{0}}^{2}(1+\frac{{c_{1}}^{2}}{{v_{0}}^{2}})]e^{2{\gamma}t}\label{70}
\end{equation}
This magnetic energy indicates that the stretching of the magnetic
field accumulates energy and gives rise to a fast dynamo, where
${\gamma}={\tau}_{0}v_{0}$. Thus one notes that the fast dynamo is
enhanced by the presence of torsion and toroidal velocity $v_{0}$
\textbf{This result is actually confirmed by Klapper and Longcope
\cite{21} assertion that the twist is influenced by the flow where
tubes are immersed. Actually the expression (\ref{69}) shows that
the stretching of the tube is enhanced by  torsion since this last
one appears in the exponential stretching. Physically this means
that the flow along the magnetic filament is stronger than any
magnetic orthonormal perturbation} and the system might be stable.
(\textbf{note to the referee: The next section is a whole new
section which is meant as an application to the above nti-dynamo
theorem}).
\section{Plasma dynamo Riemannian twisted flux tube} \textbf{This section contains a
solution of the self-induction magnetic equation , given by
\begin{equation}
{\partial}_{t}\textbf{B}=(\textbf{B}.{\nabla})\textbf{u}-(\textbf{u}.{\nabla})\textbf{B}+{\eta}{\nabla}^{2}\textbf{B}
\label{71}
\end{equation}
where the magnetic field is defined in terms of the vector potential
$\textbf{A}$ as
\begin{equation}
\textbf{B}:={\nabla}{\times}\textbf{A} \label{72}
\end{equation}
Here ${\eta}=\frac{UL}{Re_{m}}$ is the resistivity diffusion , while
the U and L are respectively, the velocity and lenght scales of the
plasma flow. Besides equation (\ref{71}) one also considers as in
laminar plasma dynamos that the plasma is a Beltrami flow given by
the equation
\begin{equation}
{\nabla}{\times}\textbf{u}={\lambda}_{B}\textbf{u} \label{73}
\end{equation}
The self-induction equation (\ref{1}) can be split into three scalar
equations if one considers the Frenet equations above. Now let us
consider briefly the Riemannian geometry of the twisted flux tube as
the heliotron, whete the Frenet torsion ${\tau}_{0}$ is taken as
constant and equal to the Frenet curvature of the magnetic heliotron
axis.  where $B_{s}$ is the toroidal component of the magnetic
field. Let us now consider the magnetic field definition in terms of
the magnetic vector potential $\textbf{A}$ as
\begin{equation}
\textbf{B}=B_{\theta}(r,{\theta})\textbf{e}_{\theta}+B_{s}(r)\textbf{t}
\label{74}
\end{equation}
The gradient operator is
\begin{equation}
{\nabla}=\textbf{t}{\partial}_{s}+\frac{1}{r}\textbf{e}_{\theta}+\textbf{e}_{r}{\partial}_{r}\label{75}
\end{equation}
Let us now consider the curvilinear coordinate relation
\begin{equation}
{\partial}_{s}\textbf{e}_{\theta}=-{\tau}_{0}sin{\theta}\textbf{t}\label{76}
\end{equation}
From all the above expressions one may finally split the
equation(\ref{1}) into the three scalar equations
\begin{equation}
\frac{{\gamma}U_{max}}{L}{{B_{s}}^{0}}={B^{0}}_{\theta}{\tau}_{0}sin{\theta}+{u}_{\theta}{\tau}_{0}sin{\theta}{\beta}
\label{77}
\end{equation}
where
\begin{equation}
{\beta}:=[\frac{{B_{s}}}{K}-\frac{{{\tau}_{0}}^{-1}{B^{0}}_{\theta}}{r}]
\label{78}
\end{equation}
\begin{equation}
\frac{{\gamma}U_{max}}{L}{{B_{\theta}}^{0}}={B^{0}}_{s}{\tau}_{0}+{\beta}{u}_{\theta}{\tau}_{0}tan{\theta}+u_{s}{\tau}_{0}csc{\theta}
\label{79}
\end{equation}
\begin{equation}
\frac{{\gamma}U_{max}}{L}{{B_{\theta}}^{0}}=-{\beta}{u}_{\theta}{\tau}_{0}tan{\theta}+i[\frac{{u}_{s}}{K}-\frac{{{\tau}_{0}}^{-1}u_{\theta}}{r}]
\label{80}
\end{equation}
Let us start solving this system by finding the conditions for a
non-dynamo mode which is obviously given by the condition
${\gamma}=0$. Substitution of this condition into the last equation
yields
\begin{equation}
-{\beta}{u}_{\theta}{\tau}_{0}tan{\theta}+i[\frac{{u}_{s}}{K}-\frac{{{\tau}_{0}}^{-1}u_{\theta}}{r}]=0
\label{81}
\end{equation}
since the second term inside the brackets is a quantity that belongs
to the real numbers and ${\beta}$ is also real, one obtains
\begin{equation}
\frac{{u}_{s}}{K}=\frac{{{\tau}_{0}}^{-1}u_{\theta}}{r} \label{82}
\end{equation}
It is also easy to note that $sin{\theta}$ vanishes when one
considers that the resistivity diffusion vanishes or that the
$Re_{m}\rightarrow{\infty}$. This vanishing of sine function implies
\begin{equation} {\tau}_{0}=\frac{2{\pi}}{a}[{\theta}_{R}-2{\pi}m]
\label{83}
\end{equation}
where m is an integer. This shows that non-dynamo modes
${\gamma}=0$, can be obtained by tuning the heliotron \cite{24}
torsion to certain non-dynamo surfaces.} \textbf{Note to the
referee: Next section is a fundamental new one since it provides us
with an example of non-stretching non-dynamo flows as in Vishik's
lemma in the case of diffusive media. \textbf{Note to the referee:
This next section gives us a further example of non-dynamos and slow
modes}\section{Non-stretching slow dynamos in twisted diffusive
Riemannian space} In this last section a new example of a more
realistic non-stretching flow in diffusive plasma medium is given.
As one shall now show this leads to a slow dynamo. Radial modes for
the existence of marginal dynamos are obtained. The approximation of
very weak torsion above is given for a thin non-stretched tube.}
\textbf{Let us now consider the diffusive self-induction equation
\begin{equation}
\frac{d}{dt}\textbf{B}={\partial}_{t}\textbf{B}+(\textbf{u}.{\nabla})\textbf{B}=-(\textbf{B}.{\nabla})\textbf{u}+{\eta}{\nabla}^{2}\textbf{B}
\label{84}
\end{equation}
when the non-stretched hypothesis is effective, the term
$(\textbf{B}.{\nabla})\textbf{u}$ has to vanish and equation
(\ref{84}) reduces to
\begin{equation}
\frac{d}{dt}\textbf{B}={\eta}{\nabla}^{2}\textbf{B} \label{85}
\end{equation}
By taking the magnetic field functional form
\begin{equation}
\textbf{B}=B_{0}(r)\textbf{t}e^{({\gamma}t+i[k_{s}s+k_{\theta}{\theta})}
\label{86}
\end{equation}
Substitution of this form into the expression (\ref{85}) yields
\begin{equation}
{\gamma}\textbf{B}={\eta}{\nabla}^{2}\textbf{B} \label{87}
\end{equation}
Since the tube is twisted one shall compute the Laplacian operator
as
\begin{equation}
{\Delta}:={\nabla}^{2}=[\textbf{e}_{r}{\partial}_{r}+\textbf{X}{\partial}_{s}]^{2}
\label{88}
\end{equation}
Here
\begin{equation}
\textbf{X}:=[\textbf{t}-\frac{{{\tau}_{0}}^{-1}}{r}\textbf{e}_{\theta}]\label{89}
\end{equation}
A simple algebraic manipulation yields
\begin{equation}
{\gamma}{B}_{0}={\eta}[{{\partial}_{r}}^{2}B_{0}+{\tau}_{0}cos{\theta}{\partial}_{r}B_{0}-[1-\frac{{{\tau}_{0}}^{-2}}{r^{2}}]
[{k_{s}}-{{\tau}_{0}}k_{\theta}]^{2}+i[{k}_{s}-{{\tau}_{0}}k_{\theta}]\frac{sin{\theta}}{r}]B_{0}]\label{90}
\end{equation}
a great deal of simplification is achieved by considering the
solenoidal character of the $\textbf{B}$ field as
\begin{equation}
\nabla.\textbf{B}=i[{k_{s}}-{{\tau}_{0}}k_{\theta}]\textbf{B}=0\label{91}
\end{equation}
which implies that
\begin{equation}
\frac{{k_{s}}}{k_{\theta}}={\tau}_{0}\label{92}
\end{equation}
Thus the poloidal wavelenght number dominates over the toroidal one
in the case of weak torsion. By substitution of this result and
considering that the a complex wave length number $k_{s}:=ik_{0}$,
into the equation (\ref{90}) yields
\begin{equation}
{{\partial}_{r}}^{2}B_{0}-{\eta}^{-1}{\gamma}B_{0}=0\label{93}
\end{equation}
Assuming that the $B_{0}(r)$ has the form
\begin{equation}
B_{0}(r)=r^{n}\label{94}
\end{equation}
where n is a real number. Thus by substituting the ansatz (\ref{94})
into the expression (\ref{93}) yields
\begin{equation}
n(n-1)r^{n-2}-{\eta}^{-1}{\gamma}r^{n}=0\label{95}
\end{equation}
By making use of the ordinary differential equation technique of
reducing both integers in (\ref{95}) to the same $r^{n}$, one has to
substitute $n\rightarrow{n+2}$ into the first term exponent of the
LHS of equation (\ref{95}). These ODE operations reduce the
algebraic equation (\ref{95}) to
\begin{equation}
[(n+2)(n+1)-{\eta}^{-1}{\gamma}]r^{n}=0\label{96}
\end{equation}
Assuming that one is never at the magnetic axis where $r=0$ one has
to solve the algebraic equation
\begin{equation}
n^{2}+3n+(2-{\eta}^{-1}{\gamma})=0\label{97}
\end{equation}
which is a second-order algebraic equation. Solution of this
equation allows us now to determine the radial n-modes which leads
to the dynamo solutions. These solutions are
\begin{equation}
n_{\pm}=-\frac{3}{2}\pm[\frac{7}{2}(1-\frac{1}{14}{\eta}^{-1}{\gamma})]=0\label{98}
\end{equation}
This solution yields the following modes for the growth of magnetic
field ${\gamma}$
\begin{equation}
{\gamma}_{+}=4(n_{+}-2){\eta}\label{99}
\end{equation}
and
\begin{equation}
{\gamma}_{-}=-4(n_{-}+5){\eta}\label{100}
\end{equation}
Both these modes are actually non-dynamos in agreement with Vishik's
lemma, since there is a direct dependence with resistivity ${\eta}$
thus when ${\eta}\rightarrow{0}$ so is ${\gamma}\rightarrow{0}$.
This is actually exactly the slow dynamo condition \cite{9}. Note
that marginal dynamo modes can be obtained by using the constraints
${\gamma}_{\pm}=0$ into respectively equations (\ref{99}) and
(\ref{100}). This yields immeadiatly the solutions corresponding to
$n_{+}=2$ and $n_{-}=-5$ as
\begin{equation}
\textbf{B}_{+}=\textbf{t}r^{2}e^{i(k_{s}s+k_{\theta}{\theta})}\label{101}
\end{equation}
and
\begin{equation}
\textbf{B}_{-}=\textbf{t}r^{-5}e^{i(k_{s}s+k_{\theta}{\theta})}\label{102}
\end{equation}
Note that the first mode is regular near to the magnetic torsioned
axis, while in the other mode the magnetic field grows in space very
fast as the magnetic field axis is approached. The role of the
curvature, either Gaussian or Riemannian, on the existence of fast
dynamos can still be better understood if one imposes their
vanishing in the Chicone-Latushkin \cite{5} expression for
${\gamma}$ in the fast dynamo obtained from geodesic flows in Anosov
spaces of negative constant Riemannian curvature
\begin{equation}
{\gamma}=\frac{1}{2}[-{\eta}(1+{\kappa}^{2})+\sqrt{{\eta}^{2}(1-{\kappa}^{2})^{2}-4{\kappa}}]\label{103}
\end{equation}
where ${\kappa}$ is the Gaussian curvature. From this expression one
notes that when this Gaussian curvature of the two-dimensional
manifold vanishes, the expression for ${\gamma}$ vanishes. This
shows that in the absence of curvature a marginal non-fast dynamo is
obtained. Therefore theorems discussed here maybe consider as
particular cases of Chiconne-Latushkin theorem \cite{5} on fast
dynamos of specific Riemannian plasma geometries. Actually in cases
investigated in this paper, stretching is deeply connected to
curvature through term $K(r,s)$ in Ricca's metric. This lead us to
conclude that the curvature is actually connected to stretching and
folding. Since in the case considering in this section, twist,
curvature and stretching are almost neglected the result seems to be
physically correct.
\section{Conclusions} \textbf{Vishik's idea that the non-stretched
flows cannot be fast dynamos, is tested once more here in several
examples. All these examples involve either stretching or
non-stretching of non-dynamo modes, proving in this way, that even
stretching flux tubes like the plasma heliotrons for example, may
not provide a dynamo action. This constitute a new framework to
discuss anti-dynamo theorems}. When twist or Frenet torsion is
small, slow dynamos are shown to be present in these astrophysical
loops. By following the analogy proposed by Friedlander and Vishik
in dynamo theory between vorticity equations and dynamo equations
and considering exponential stretching one shows that flux tube
dynamos are in complete agreement with Ricca's original Riemannian
magnetic flux tube model. STF Zeldovich-Vainshtein  fast dynamo
generation method ,is not the only Riemannian method that can be
applied as in Arnold's cat map, but other conformal fast kinematic
dynamo models as the Riemannian one, \textbf{can also be useful as
new dynamo action generation method}. Small scale dynamos in
Riemannian spaces can therefore be very useful for our understanding
of more large scale astrophysical dynamos. Other applications of
plasma filaments such as stretch-twist and fold fractal dynamo
mechanism, which are approximated Riemannian metrics have been
recently put forward by Vainshtein et al \cite{8}. Finally one has
shown that the Vishik's result is strongly enhanced, and no fast
dynamo and sometimes even slow dynamo is found in non-stretching
flows. Thus decay of the magnetic field in non-stretched filaments
in presence of magnetic diffusion and the Frenet torsion is found.
As considered by Arnold \cite{16} and Vainshtein et al \cite{2} no
fast dynamos exists in \textbf{2D} Euclidean space and a fast dynamo
was obtained in 3D curved Riemannian space by Arnold \cite{16} where
stretching in some directions is compensated by compression in the
other, \textbf{in somewhat artificial uniform stretching}. In this
paper a fast filamentary dynamo was obtained in Euclidean 3D space.
Note that when torsion vanishes in the last section a marginal
dynamo is obtained. \textbf{This} result, a sort of filamentary
antidynamo theorem , has been recently obtained \cite{25} in the
non-holonomic frame. \textbf{A more realistic diffusive media is
used to obtain an example of Riemannian plasma tubes, where no fast
dynamo action is obtained. One notes that is very easy to
generalizes the results obtained here to the non-uniformly
stretching flows by simply replacing the Ricca's Riemannian metric
by
$d{s^{2}}_{0}=H(s-s_{0})[dr^{2}+r^{2}d{{\theta}_{R}}^{2}]+K^{2}(r,s)ds^{2}$,
where H is the stept Heaviside function. This new Riemann flux
"fracture tube" metric leads to a fast dynamo flow where the flows
are proportional to the Dirac delta function similar to chaotic
periodic dynamo flows investigated by Finn and Ott \cite{26}. The
detailed computations and further discussions may appear
elsewhere.}}
\section{Acknowledgements} I am deeply greatful to Renzo Ricca
and Yuri Latushkin for their extremely kind attention to our work.
Thanks are also due to I thank financial supports from Universidade
do Estado do Rio de Janeiro (UERJ) and CNPq (Brazilian Ministry of
Science and Technology).


\newpage
\begin{thebibliography}{26}
  \bibitem{1} M. Vishik, Izv Acad Sci USSR, Phys Solid Earth 24(1), 173(1988).
  \bibitem{2} S. Cowling , Magnetohydrodynamics (1964) Oxford.
  S.I. Vainshtein, Ya B Zeldovich, Sov Phys Usp 15
  ,159 (1972). S.Vainshtein, A. Bykov and I.N. Toptygin,
  Turbulence, Current Sheets and Shock waves in Cosmic Plasmas, Gordon
  \bibitem{3} S. Friedlander, M. Vishik, Chaos 1(2),198 (1991).
  \bibitem{4} D.V. Anosov, Geodesic Flows on Compact Riemannian
  Manifolds of Negative Curvature, (Steklov Mathematical
  Institute, USSR) vol.90 (1967).
  \bibitem{5} C. Chicone and Yu Latushkin, Evolution
  Semigroups in Dynamical systems and differential equations, American
  Mathematical Society, AMS-(1999). C. Chicone and Yu Latushkin, Proc of the American
 Mathematical Society 125,N. 11,3391 (1997).
 \bibitem{6} L. C. Garcia de Andrade, Physics of Plasmas 14 (2007).
 \bibitem{7} M. Vishik, Izv Acad Sci USSR, Phys Solid Earth 24(1),173(1988).
 \bibitem{8} S.Vainshtein, A. Bykov and I.N. Toptygin,
  Turbulence, Current Sheets and Shock waves in Cosmic Plasmas, (1998) Gordon
  \bibitem{9} S. Childress, A. Gilbert, Stretch, Twist
  and Fold: The Fast Dynamo (1996),Springer, Berlin.
  \textbf{\bibitem{10} I Klapper and L Young , Comm Math Phys. \textbf{173} 623
  (1995).}
  \bibitem{11} S Molchanov, Russian mathematicals surveys (1978).
  \bibitem{12} A M Soward, Geophys Astrophys Fluid Dyn \textbf{53},81 (1990).
  \textbf{\bibitem{13} A. B. Mikhailovskii, \textbf{Instabilities in a confined Plasma}, IOP (1998).
  \bibitem{14} R. Ricca, Solar Physics \textbf{172} (1997),241.
  \bibitem{15} L C Garcia de Andrade, Phys Plasmas \textbf{13}
  (2006). L. C. Garcia de Andrade, Non-holonomic dynamo filaments
  as Arnold´s map in Riemannian space, Astronomical notes (2008) in
  press.
  \bibitem{16} V. Arnold, Ya B. Zeldovich, A. Ruzmaikin and D.D. Sokoloff, JETP 81 ,n. 6,
  2052 (1981). V. Arnold, Appl Math and Mech \textbf{36}, 236 (1972).
  \bibitem{17} Z Wang, V Pariev, C Barnes and D Barnes, Phys Plasmas \textbf{9},5 (2002).
  \bibitem{18} P M Bellan, Spheromaks: A practical application of
  MHD Dynamos and Plasma self-organization, (2002) Imperial College
  Press.
  \bibitem{19} A Jacobson, Phys Fluids \textbf{27} (1),7 (1984).
  \bibitem{20} M. Berger and C Prior, J Phys A \textbf{39}, 8321 (2006). P Newton, The N-Vortex problem (2001) Springer.
  \bibitem{21} M.C. Lopez Fuentes, P. Demoulin, C.H.
   Mandrini, A.A. Pevtsov and L. van Driel-Gesztelyi, Astr and
   Astrophys. \textbf{397}, 305 (2003).
  \bibitem{22} M. Nu\~{n}ez, J. Phys. A:Math and Gen. \textbf{36}, 8903 (2003).
  \bibitem{23} I Klapper and D Longcope, Evolution Equations of Thin
  Twisted Flux Tubes, in "Workshop on Stellar Dynamos ASP conference series, (1999) M. Nunez and Ferriz-Mas eds.
  \bibitem{24} M. Wakatani, Stellarator and Heliotron Devices,(1998)
  Oxford University Press.
  \bibitem{25} L. C. Garcia de Andrade, Non-holonomic dynamo filaments
  as Arnold´s map in Riemannian space, Astronomical notes (2008) in
  press.
  \bibitem{26} J M Finn and E Ott, Phys Fluids \textbf{31} (1983).}

  \end{thebibliography}
  \end{document}